\documentclass[conference]{IEEEtran}
\IEEEoverridecommandlockouts
\usepackage{cite}
\usepackage{amsmath,amssymb,amsfonts,subcaption}
\usepackage{algorithmic}
\usepackage{graphicx}
\usepackage{textcomp}
\usepackage{xcolor}
\definecolor{matlabblue}{RGB}{0, 114, 188}
\definecolor{matorange}{RGB}{217, 83, 25}
\definecolor{matyellow}{RGB}{237, 177, 32}
\definecolor{matpurple}{RGB}{127, 0, 255}
\definecolor{matgreen}{RGB}{119, 170, 48}
\definecolor{matlightblue}{RGB}{77, 190, 238}
\definecolor{matred}{RGB}{162, 20, 47}
\def\BibTeX{{\rm B\kern-.05em{\sc i\kern-.025em b}\kern-.08em
    T\kern-.1667em\lower.7ex\hbox{E}\kern-.125emX}}
\begin{document}

\title{Enhancing LMMSE Performance with Modest Complexity Increase via Neural Network Equalizers}

\author{\IEEEauthorblockN{Vadim Rozenfeld}
\IEEEauthorblockA{
\textit{Tel-Aviv University, Israel.}\\
vadimroz31@gmail.com}
\and
\IEEEauthorblockN{Dan Raphaeli}
\IEEEauthorblockA{\textit{Tel-Aviv University, Israel.}\\
danir@eng.tau.ac.il}
\and
\IEEEauthorblockN{Oded Bialer}
\IEEEauthorblockA{
\textit{General Motors.}\\
oded.bialer8@gmail.com}
}
\maketitle
\begin{abstract}
The BCJR algorithm is renowned for its optimal equalization, minimizing bit error rate (BER) over intersymbol interference (ISI) channels. However, its complexity grows exponentially with the channel memory, posing a significant computational burden. In contrast, the linear minimum mean square error (LMMSE) equalizer offers a notably simpler solution, albeit with reduced performance compared to the BCJR.
Recently, Neural Network (NN) based equalizers have emerged as promising alternatives. Trained to map observations to the original transmitted symbols, these NNs demonstrate performance similar to the BCJR algorithm. However, they often entail a high number of learnable parameters, resulting in complexities comparable to or even larger than the BCJR. This paper explores the potential of NN-based equalization with a reduced number of learnable parameters and low complexity. We introduce a NN equalizer with complexity comparable to LMMSE, surpassing LMMSE performance and achieving a modest performance gap from the BCJR equalizer.
A significant challenge with NNs featuring a limited parameter count is their susceptibility to converging to local minima, leading to suboptimal performance. To address this challenge, we propose a novel NN equalizer architecture with a unique initialization approach based on LMMSE. This innovative method effectively overcomes optimization challenges and enhances LMMSE performance, applicable both with and without turbo decoding.

\end{abstract}


\section{Introduction}
Reliable data transmission over an ISI channel is vital for modern communication systems, where multiple symbols interfere, distorting the received signal and complicating decoding. At the receiver, an equalizer is deployed to mitigate ISI effects and reconstruct the original signal. With known channel state information (CSI) to the receiver, equalizer and decoder can iteratively work together in a turbo equalization \cite{b1} manner, where feedback information from the decoder is looped back to the equalizer.

The optimal equalization technique for BER minimization is nonlinear, e.g., BER-optimal BCJR algorithm \cite{b2}, maximizing the \textit{a posteriori} probability (MAP) estimation of the data. However, full-state BCJR entails exponential computational complexity growth with channel memory. Reduced-state BCJR methods \cite{b3, b4, b5} address complexity concerns, but may exhibit inferior performance compared to full-state BCJR. Additionally, the complexity of reduced-state alternatives still grows exponentially with channel memory.

An LMMSE equalizer \cite{b6}, provides a computationally efficient alternative, albeit with a trade-off in terms of BER performance. In turbo equalization, BCJR typically outperforms LMMSE due to its ability to generate more reliable log-likelihood ratio (LLR) outputs, regardless of the signal-to-noise ratio (SNR) and the number of iterations employed.

Contrary to the model-based algorithms reviewed above, data-driven models, particularly NNs, are alternative approach for performing equalization tasks. Unlike traditional model-based algorithms, data-driven models are trained to directly map observations (data) to the originally transmitted symbols. 
In fact, NNs are capable to extract relevant features from observations themselves, and therefore a data-driven equalizer does not require to be explicitly fed with CSI, assumed trained on sufficiently large dataset. As an example, N-BCJR \cite{b7} is recurrent neural network (RNN) based decoder for sequential codes with close to optimal performance on the AWGN channel. The work \cite{b8} proposes a joint equalizer and decoder, each of which is implemented via NN, to achieve equalization and decoding processes without requiring knowledge of the CSI. The work \cite{b9} demonstrates blind equalization via variational auto-encoder (VAE).
Examples of hybrid data-driven/model-driven receivers proposed for symbol detection tasks, both with and without CSI available to the receiver, include BCJRNet \cite{b10} and ViterbiNet \cite{b11}.

While many of the aforementioned models demonstrate performance within a small gap of conventional algorithms, this often comes at the prohibitive cost of a high number of learnable parameters and operations carried out per received symbol per iteration.
In fact, BCJRNet and ViterbiNet already have thousands of learnable parameters, far exceeding computational load of full-state BCJR. In MetNet \cite{b12}, the complexity of ViterbiNet is further reduced, although it remains three times higher than that of the full-state BCJR.

Additionally, the models reviewed above are not trained to integrate seamlessly with standard decoders or to perform optimally in turbo equalization.

In this paper, we investigate the performance potential of NN-based equalization using a limited number of learnable parameters. The benefit of such NN lies in its reduced computational complexity and fast running time. We design an NN equalizer with complexity comparable to LMMSE, which surpasses LMMSE and achieves a modest performance gap from the MAP equalizer. A significant challenge with NNs having a small parameter count is their susceptibility to converging to local minima, resulting in sub-optimal performance. To address this, we propose a novel NN equalizer architecture built upon LMMSE, which enhances its capabilities, applicable both with and without turbo decoding. This approach allows us to construct a compact NN that can be initialized with LMMSE to mitigate optimization issues.

The main contributions of the paper are as follows:
\begin{enumerate}
\item Highlighting the difficulty in optimizing neural network equalizers with limited parameters, which tend to settle into sub-optimal local minima, failing to surpass LMMSE performance.
\item Introducing a NN-based equalizer design that addresses the optimization challenge and exceeds LMMSE performance while maintaining similar complexity levels.
\item Showcasing that NN-based equalizers offer a low-complexity solution for narrowing the BER gap between LMMSE and MAP, both with and without decoding.
\end{enumerate}

\section{System Model and Equalization Algorithms}
We consider a system with an information source producing independent and uniformly distributed (IUD) data bits \(\mathbf{b}\triangleq[b_0,b_1,\cdots, b_{K-1}]\). For integrity reasons, a block is protected by LDPC code, yielding a code-word $\mathbf{c}\triangleq[c_0, c_1, \ldots, c_{U-1}]$. It is then partitioned into $Q$ blocks, each with $q=\log_2 M$ bits, $\mathbf{c}=[\mathbf{c}_0, \ldots, \mathbf{c}_{Q-1}]$, and mapped to M-ary PAM symbols $\mathbf{x}\triangleq[x_0, x_1, \ldots, x_{Q-1}]^T$ using a binary Gray-code function. 
The AWGN ISI channel model is given by:
\begin{equation}
\label{eq:ISI_model}
    z_n =h_0x_n+ \sum\limits_{i=1}^{M_h}h_ix_{n-i} + w_n, \quad n=0, 1,\ldots, Q-1,
\end{equation}
where the second term is ISI. We assume $x_n,w_n \text{ and } z_n$ are zero outside $\{0, 1, \ldots, Q-1\}$. Here, \(\mathbf{h}=[h_0, h_1,\cdots, h_{M_h}]\) is the invariant and known to the receiver channel impulse response (CIR), \(\mathbf{z}=[z_0,z_1,\cdots, z_{Q-1}]^T\) be the receiver input and \(w_n\sim\mathcal{N}(0, \sigma^2_w)\) are noise samples, all are real-valued. 
On the receiver side a channel equalizer is operating to counteract the distortion caused by the ISI channel \(\mathbf{h}\) and a decoder is then recovering the original data. As both equalizer and decoder operate on same set of received data, it is quite beneficial to recover the data bits via turbo equalization, which involves iteratively exchanging information between the soft-in soft-out (SISO) equalizer and a SISO decoder. Namely, the equalizer associates an LLR estimation to code bit $c_n$ using:
\begin{equation}
    L(c_n) \triangleq \ln \left(\frac{P(c_n=0 | \mathbf{z})}{P(c_n=1 | \mathbf{z})}\right), \quad n=0,1,\ldots U-1,
\end{equation}
or the corresponding soft-bit \cite{b15}:
\begin{equation}
    \label{eq:soft_bits}
    \Lambda(c_n)=\tanh{(L(c_n)/2)}.
\end{equation}
While common choices for equalizers include MAP and LMMSE, our interest lies in an NN-based alternative. 

\subsection{LMMSE and MAP Equalizers}
LMMSE estimates $\hat{x}_n$ by minimizing MSE $\mathbb{E}(|x_n-\hat{x}_n|^2)$. To this end, it processes a frame of observations with a length of \(N=N_1+N_2+1\) at time \(n\) which described by:
\begin{equation}
    \mathbf{z}_n\triangleq[z_{n-N_2},\dots,z_n, \dots,z_{n+N_1}]^T 
    \label{eq:input_window}
\end{equation}
 to compute the estimation \(\hat{x}_n\) of \(x_n\) by:
\begin{equation}
    \hat{x}_n=\mathbb{E}(x_n)+\mathrm{Cov}(x_n,\mathbf{z}_n)\mathrm{Cov}(\mathbf{z}_n, \mathbf{z}_n)^{-1}(\mathbf{z}_n-\mathbb{E}(\mathbf{z}_n)).
    \label{eq:lmmse}
\end{equation}
For initial equalization step with no \textit{a priori} information available, \cite{b6} suggests \(\mathbf{f}=[f_{N_2},\ldots,f_0, \dots,f_{-N_1}]^T\) for each time step is set to:
\begin{equation}
    \mathbf{f}\triangleq(\sigma^2_w\mathbf{I}_N+\mathbf{H H^T})^{-1}\mathbf{h}_n,
    \label{eq:lmmse_filter}
\end{equation}
with $f_0$ as center tap of $\mathbf{f}$.
Here, \(\mathbf{H}\) is $N\times(N+M_h)$ channel convolution matrix with $1\times (N+M_h)$ coefficients vector $[0, \ldots, 0, h_{M_h}, h_{M_h-1}, \ldots, h_{0}, 0, \ldots, 0]$, \(\mathbf{h}_n\) is the (\(M_h+N_2)\)th column of \(\mathbf{H}\) and \(\mathbf{I}_N\) is \(N\!\times\!N \) identity matrix. And so, the LMMSE estimator of \(\hat{x}_n=\mathbb{E}(x_n \lvert \mathbf{z}_n)\) is expressed as:
\begin{equation}
    \hat{x}_n = \sum\limits_{i=-N_1}^{N_2} f_{i}z_{n-i}=\mathbf{f}^T\mathbf{z}_n, \quad n=0,1,\ldots,Q-1.
    \label{eq:lmmse_equalizer}
\end{equation}

The MAP symbol detection algorithm is aimed to minimize symbol error probability $P(\hat{x}_n \ne x_n)$ by setting $\hat{x}_n$ to:
\begin{equation}
    \hat{x}_n=\operatorname*{argmax}_{x\in \mathcal{B}} \; P(x_n=x|\mathbf{z}), \quad n=0, 1, \ldots,Q-1.
\end{equation}
This can be efficiently achieved with BCJR if the alphabet size \(|\mathcal{B}|\) and/or the channel memory \(M_h\) are small enough, as BCJR's complexity is \(\mathcal{O}(|\mathcal{B}|^{M_h})\).
\begin{figure}[t]
    \centering
    \includegraphics[width=0.7\linewidth]{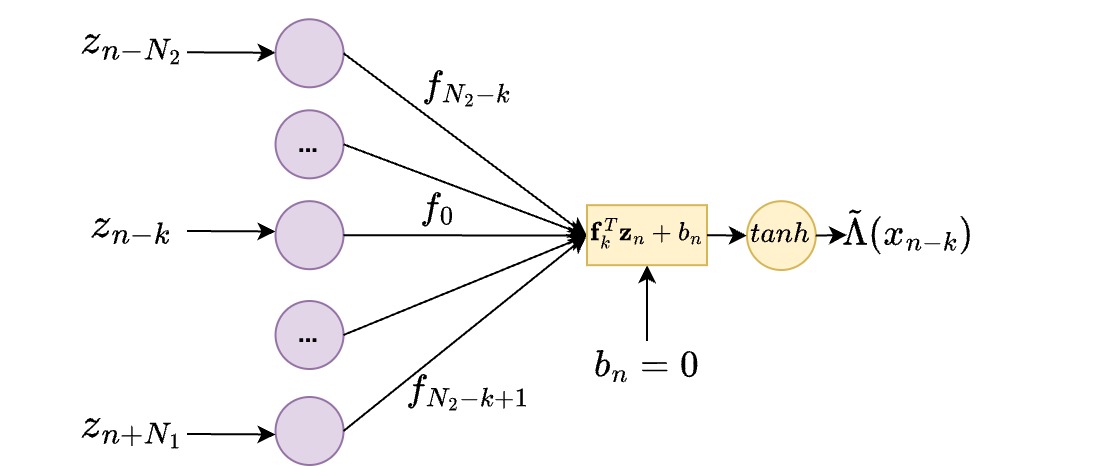} 
    \caption{An approximated soft-bit estimator for $x_{n-k}$, utilizes $k\!>\!0$ time units shifted LMMSE taps, to initialize the weights.} 
    \label{fig:single_neuron} 
    \vspace{-1em}
\end{figure}

\begin{figure}[t]
    \centering
    \includegraphics[width=0.7\linewidth]{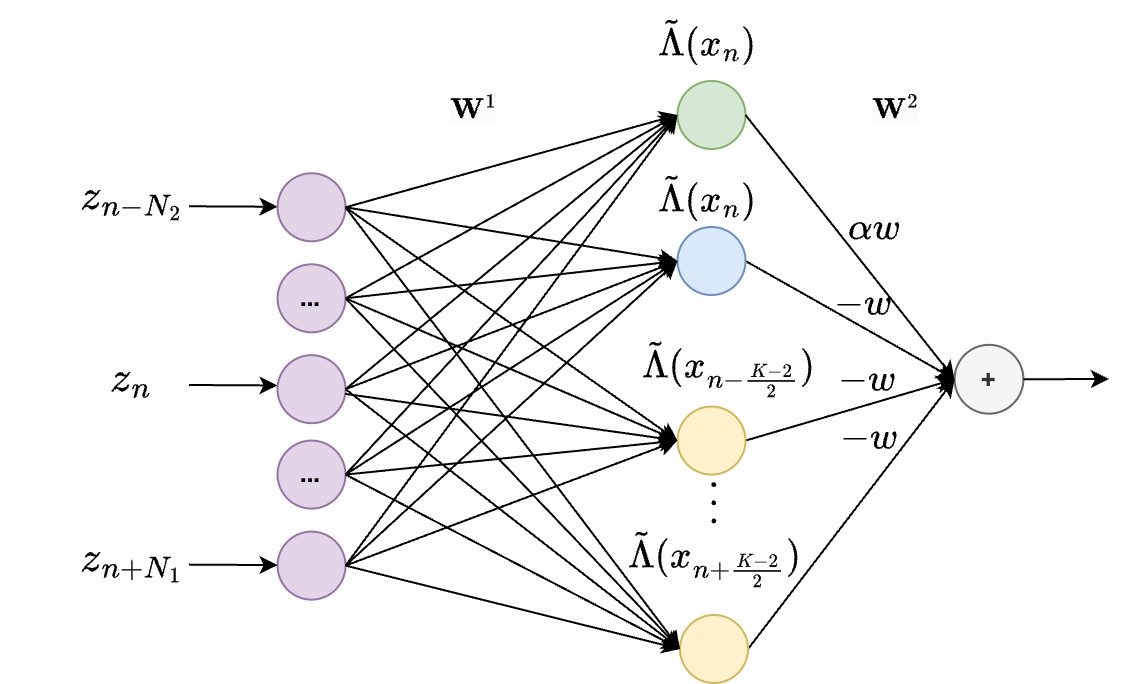}
    \caption{$K$-EqzNet: An equalizer for symbol $x_n$, processes three types of soft-bits: green for $x_n$, blue for $-x_n$, and yellow for symbols in range $\{x_{n-k}, \ldots, x_{n+k}\} \text{ for } k=(K-2)/2$, responsible for the ISI at $z_n$. Soft-bits are scaled by weights $\mathbf{W}^2$ to produce LLRs. The initial weights $\mathbf{W}^1$ from (\ref{eq:k_time_unis_shifts_soft_bits}) are omitted here. All parameters are updated during training.} 
    \label{fig:multi_neurons}
    \vspace{-1em}
\end{figure}

\begin{figure}[t]
    \centering
    \includegraphics[width=0.9\linewidth]{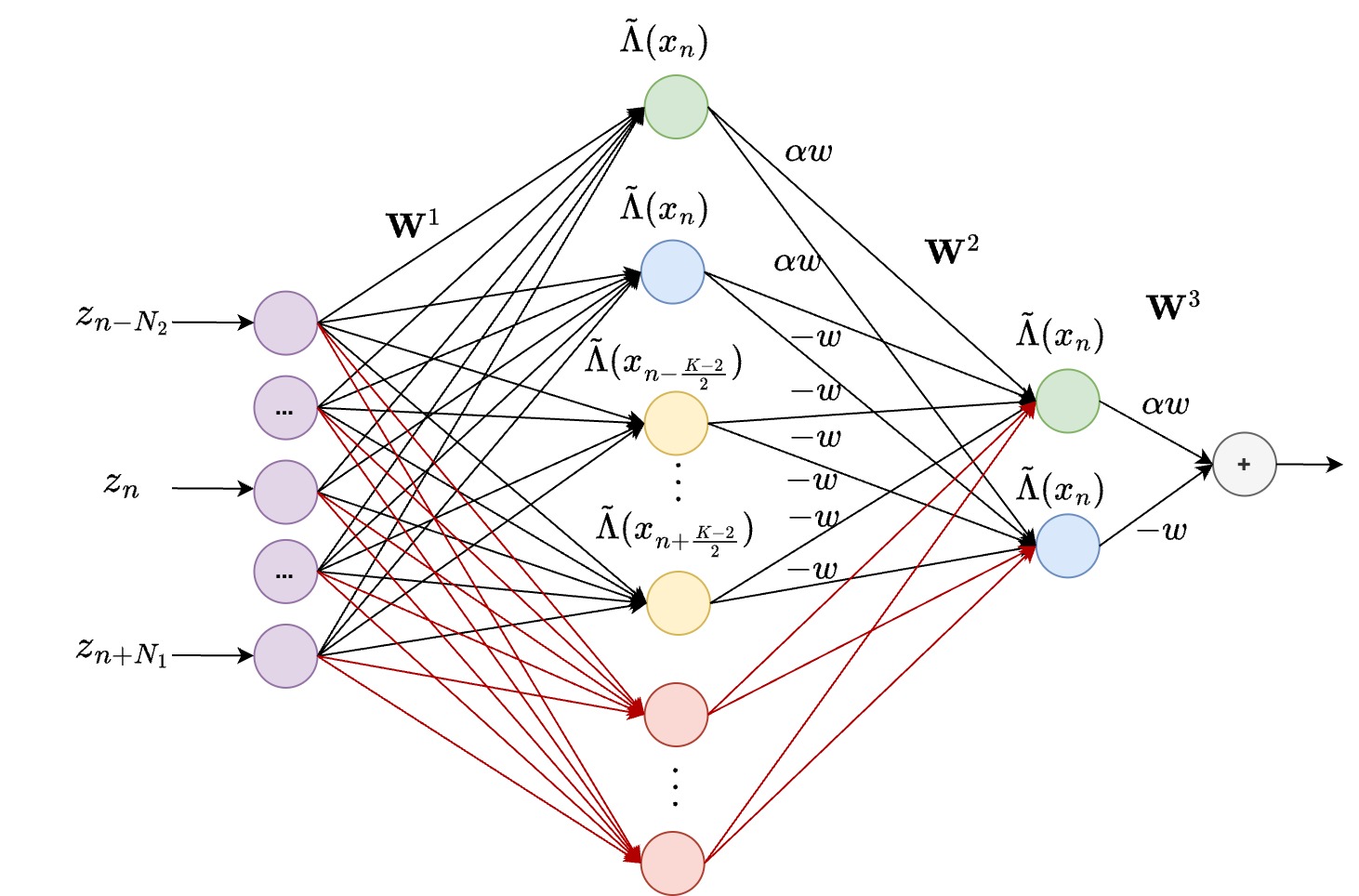}
    \caption{ ($K+L,2$)-EqzNet: Extending $K$-EqzNet by integrating a pretrained $L$-EqzNet, distinguished by its red color. The outputs from both networks are fully connected to a hidden-layer, producing LLRs for $x_n$ and $-x_n$. All parameters, including those from $L$-EqzNet, are updated during training.} 
    \label{fig:finaml_model} 
    \vspace{-1em}
\end{figure}

\section{The Proposed Neural Network Equalizer}
\label{sec:proposed_nn}
Our proposed equalizer is based on a fully-connected neural network (FC-NN) with at least one hidden layer containing non-linearity (e.g., $\tanh$ as in \eqref{eq:soft_bits}) to capture intricate patterns and complex features from the raw input data. This network takes as its input a frame \(\mathbf{z}_n\) as expressed in (\ref{eq:input_window}). The central observation $z_n$ within this frame corresponds to the sent symbol $x_n$. It produces as output an LLR estimation, \(L(x_n)\), of the originally sent symbol $x_n \in \{-1, +1\}$.
The length of \(\mathbf{z}_n\), typically longer than channel memory \(M_h\), is an optimized hyperparameter. 
The FC-NN's layer \(\ell\) takes input vector \(\mathbf{y}^{\ell-1}\) and outputs vector \(\mathbf{y}^{\ell}\) through:
\begin{equation}
    \label{eq:inner_node_fcnn}
    \mathbf{y}^{\ell} = \phi(\mathbf{W^{\ell}}\cdot \mathbf{y}^{\ell-1}+\mathbf{b}^{\ell}), \quad \ell=1,2 \ldots, L,
\end{equation}
where $\mathbf{y}^0$ is the input layer and $\mathbf{y}^{L}$ is the output layer. Matrix \(\mathbf{W^{\ell}}\) is the weights learned at $\ell$-th layer, vector \(\mathbf{b}^{\ell}\) is the bias term and \(\phi\) is the non-linearity (activation function) applied element-wise.
Let \( \boldsymbol{\theta} = (\mathbf{W}^{1}, \mathbf{W}^{2}, \ldots, \mathbf{W}^{L}, \mathbf{b}^{1}, \mathbf{b}^{2}, \ldots, \mathbf{b}^{L})\) be the set of all parameters in a NN with \( L \) layers and let \(f_{\boldsymbol{{\theta}}}(\mathbf{z}_n)\) correspond to it's output for input \(\mathbf{z}_n\).
Also, let \(\mathcal{D} = \{(\mathbf{z}_n,LLR^{MAP}(x_n))\}_{n=1}^{m}\) be a dataset with \(m\) training examples, where \(\boldsymbol{z}_n\) are the frames and \(LLR^{MAP}(x_n)\) be the corresponding MAP-equalizer LLRs for symbol \(x_n\) in each frame. The FC-NN is trained to minimize the loss function:
\begin{equation}
\label{eq:loss}
\mathcal{L}_{\mathcal{D}}(\boldsymbol{\theta}) = \frac{1}{m} \sum_{n=1}^{m} \| f_{\boldsymbol{{\theta}}}(\mathbf{z}_n) - LLR^{MAP}(x_n) \|^2.
\end{equation}
That is, the FC-NN is optimized to produce soft-outputs to guide the SISO decoder, with the aim of attaining LLRs that match those generated by a MAP equalizer.

Overparametrized Deep NNs (DNNs) have demonstrated remarkable success when trained using gradient-based methods, despite the non-convex nature of the associated optimization problem (e.g., our loss function at (\ref{eq:loss})). In fact, overparametrization typically ensures that the local minima achieved are either global minima or nearly as good as the global minimum. Nevertheless, our aim is to develop an NN-based equalizer that not only maintains a complexity comparable to that of the LMMSE but also operates effectively with small number of parameters. Given this constraint, training becomes challenging as poor local minima, existing due to the non-convexity of (\ref{eq:loss}), can fail the gradient-based method \cite{b13}, as further demonstrated in Section \ref{sec:experiments and results}.

In the next subsection, we present our method for overcoming the convergence issue by initializing weights based on LMMSE filters and employing a progressive learning approach, resulting in performance enhancement over LMMSE.

\subsection{Initialization and Progressive Learning}
\label{subsec:2_nn_model}
We now explore the architecture, initialization, and progressive learning aspects of our model in detail. The cornerstone of our approach involves initializing the first hidden-layer of the FC-NN such that each neuron outputs a soft-bit estimation (\ref{eq:soft_bits}) of the symbol $x_n$ or symbols positioned $k$ time units away, namely, $x_{n-k}$ and $x_{n+k}$. With $x_n\in\{-1,+1\}$, we utilize \cite{b6} to have an LLR estimate for symbol $\hat{x}_n$ derived at \eqref{eq:lmmse_equalizer}:
\begin{equation}
    \label{eq:lle_lmmse}
    L(x_n)=(2\mathbf{f}^T\mathbf{z}_n)/({1-\mathbf{h}_n^T\mathbf{f}}),
\end{equation}
which is then substituted into (\ref{eq:soft_bits}), resulting in the following: 
\begin{equation}
\label{eq:single_node_in_FCNN}
    \Lambda(x_n)=\tanh\bigg( \frac{\mathbf{f}^T\mathbf{z}_n}{1-(\mathbf{h}_n)^T\mathbf{f}} \bigg)=\tanh\bigg(\frac{\mathbf{f}^T\mathbf{z}_n}{C}\bigg).
\end{equation}
To align the soft-bit expression above with the standard form of a neuron in FC-NN (\ref{eq:inner_node_fcnn}), we simplify (\ref{eq:single_node_in_FCNN}) by setting $C=1$. This modification primarily affects the reliability (magnitude) of soft-bit \(|\Lambda(x_n)|\) without changing its sign. 
Accordingly, from (\ref{eq:inner_node_fcnn}) and relaxed version of (\ref{eq:single_node_in_FCNN}), we define an approximated soft-bit estimation of $x_{n-k}$ as the output of a neuron, as illustrated at Fig. \ref{fig:single_neuron}, and expressed by:
\begin{equation}
\label{eq:k_time_unis_shifts_soft_bits}
    \Tilde{\Lambda}(x_{n-k}) \triangleq \tanh(\mathbf{f}_{k}^T \mathbf{z}_n+b_n),
    \quad k = \ldots, -1, 0, 1, \ldots
\end{equation}
with $b_n=0$. Here, \(\mathbf{f}_k\) corresponds to the LMMSE filter $\mathbf{f}$ \eqref{eq:lmmse_filter} cyclically shifted by \(k\) time units. 
This yields a good approximation for soft-bit estimations of transmitted symbol $x_{n-k}$ over sufficiently long frames.
We now leverage this insight to introduce our first equalizer that estimates the LLR of $x_n$ given observations frame $\mathbf{z}_n$, through scaled soft-bits:
\begin{align}
\label{eq:K_block_output}
f_{\boldsymbol{{\theta}}}&^{\text{$K$-EqzNet}}(\mathbf{z}_n) = 
w_{2, 1}\Tilde{\Lambda}({x}_n) - 
w_{2, 2}\Tilde{\Lambda}({x}_n) \\ &-
\sum_{\substack{k=-(K-2)/2 \\ k\ne 0}}^{(K-2)/2}w_{2, k+K}\Tilde{\Lambda}({x}_{n-k}) \text{ for $K$ even and $K\ge2$}. \nonumber
\end{align}
Noticeably, it has three terms as follows:
\begin{itemize}
    \item[$\bullet$] $w_{2, 1}\Tilde{\Lambda}({x}_n)$: The LLR estimation for true symbol $x_n$.
    \item[$\bullet$] $-w_{2, 2}\Tilde{\Lambda}({x}_n)$: The LLR estimation assuming the true symbol is $-x_n$, thereby attempting to correct erroneous estimations made by the first term.
    \item[$\bullet$] $\sum_{{k=-(K-2)/2, k\ne 0}}^{(K-2)/2}w_{2, k+K}\Tilde{\Lambda}({x}_{n-k})$:
This is an ISI estimation for observation \( z_n \). It is obtained through LLR estimations of transmitted symbols $\{x_{n-k},\ldots,x_{n-1}, x_{n+1}\ldots,x_{n+k}\} \text{ for } k=(K-2)/2$, and \( K \) is an hyperparameter. The ISI estimation is then subtracted from the LLR estimate of symbol \( x_n \).
    \end{itemize}
Clearly, \eqref{eq:K_block_output} defines a two-layer FC-NN depicted at Fig. \ref{fig:multi_neurons}, where the neurons at the first layer are represented by $\Tilde{\Lambda}({x}_{n-k})$. That is, first layer weight matrix is set by $\mathbf{W}^1=[\mathbf{f}^T, \mathbf{f}^T, \mathbf{f}^T_{-(K-2)/2}, \ldots, \mathbf{f}^T_{(K-2)/2}]$ and $\mathbf{b}^1=\mathbf{0}$.
Next, we initialize the weights of the second layer as follows:
$\mathbf{W}^2 = [w_{2, 1}, -w_{2, 2}, \ldots, -w_{2, K}]^T = [\alpha w, -w, \ldots, -w]^T$, for hyperparameters \(\alpha > 1\) and \(w > 0\). Here, $w_{2, 1}$ is the weight associated with the soft-bit $\tilde{\Lambda}(x_n)$. Since \(\Tilde{\Lambda}({x}_n)\) is a neuron providing the true soft-bit for the desired symbol \(x_n\) at (\ref{eq:K_block_output}), we initially weight its output \(\alpha\) times higher than others. 

Upon FC-NN is initialized as detailed above, its parameters, $\boldsymbol{\theta}^{*}=[\mathbf{W}^1,\mathbf{W}^2,\mathbf{b}^1]$, are adapted via gradient-based method to minimize \eqref{eq:loss}. 
We refer to this equalizer as $K$-EqzNet, with $K$ representing the number of neurons in its first hidden layer. It takes as an input observations frame $\mathbf{z}_n$ and outputs the LLR of transmitted symbol $x_n$ located at the center of $\mathbf{z}_n$.

In Section \ref{sec:complexity}, we show that the computational complexity of $K$-EqzNet is $K\cdot\mathcal{O}(LMMSE)$, where $\mathcal{O}(LMMSE)$ denotes the complexity of applying LMMSE to input frame $\mathbf{z}_n$. This underscores the need for $K$-EqzNet designs with $K\!<\!10$ to ensure a computational complexity on the same order of magnitude as the LMMSE filter, while also achieving enhanced BER performance. Simulations suggest that $K$ can progressively increase up to $K = M_h + 2$ before any saturation in performance becomes apparent.
To address the latter obstacle, we further extend first hidden-layer by incorporating pretrained $L$-EqzNet block, for hyperparameter $L$. The new model, termed ($K+L$)-EqzNet, takes the form:
\begin{equation}
    f_{\boldsymbol{{\theta}}}^{\text{($K+L$)-EqzNet}}(\mathbf{z}_n) = f_{\boldsymbol{{\theta}_1}}^{\text{$K$-EqzNet}}(\mathbf{z}_n) + f_{\boldsymbol{{\theta}_2}}^{\text{$L$-EqzNet}}(\mathbf{z}_n),
\end{equation}
with \(\boldsymbol{\theta}=\boldsymbol{\theta}_1 \cup \boldsymbol{\theta}_2\). This model is then trained to find new set of parameters $\boldsymbol{\theta}$ that minimizes loss function (\ref{eq:loss}). The rationale behind this approach is to leverage the benefits of an already enhanced equalizer, $L$-EqzNet, during training, resulting in improved optimization convergence.

The latter approach yields an improvement in performance only up to a certain limit before reaching saturation once more. To further enhance performance, we fully-connect the outputs of $f_{\boldsymbol{\theta_1}}^{\text{$K$-EqzNet}}(\mathbf{z}_n)$ and $f_{\boldsymbol{\theta_2}}^{\text{$L$-EqzNet}}(\mathbf{z}_n)$ into a new hidden-layer comprising two neurons, as illustrated at Fig.\ref{fig:finaml_model}. Conceptually, the additional hidden-layer is designed so that one neuron provides a soft-bit estimation for $x_n$, and the other for $-x_n$. Their scaled outputs are aggregated to estimate the LLR of $x_n$. The architecture of this additional layer is found at (\ref{eq:K_block_output}) for $K=2$, but in this instance, we do not initialize its weights with LMMSE filter taps.
This enhanced model is termed ($K+L,2$)-EqzNet. 

To summarize, we present three FC-NN models: the fundamental $K$-EqzNet, and its extended architectures ($K+L$)-EqzNet and ($K+L,2$)-EqzNet. These variations build upon $K$-EqzNet, gradually enhancing its performance with linearly increase in complexity, as elaborated in Section \ref{sec:experiments and results}.

\subsection{Fully-Connected Neural Network in Turbo Equalization}
To successfully integrate a SISO NN-based equalizer into a turbo equalization process, the equalizer must effectively handle and utilize soft information fed back from the SISO decoder at each iteration. It should then output high-quality LLRs, ideally comparable to those produced by a MAP equalizer. These LLRs are critical as they guide the decoder, influencing the overall performance of the equalization process. Failure to provide accurate LLRs can compromise the effectiveness of the decoder.

One method to ensure the NN-based equalizer meets these demands involves training a specific NN-based equalizer for each iteration (e.g., N-Turbo\cite{b7}). This is crucial because, with each iteration, ISI is progressively mitigated, reflecting a reduced noise level. Consequently, each NN-based equalizer must be tailored to match the changing noise conditions at each iteration. Given the multiple iterations involved in turbo equalization, this approach would require a corresponding number of distinct NN-based equalizers, making it impractical due to complexity and resource demands. 

An alternative strategy is to train a single NN-based equalizer to be invoked at all iterations. Nevertheless, to adapt to the continuously changing noise levels at every iteration, the complexity of this single equalizer would inevitably increase, making it prohibitive in our context.

Our chosen strategy involves employing an NN-based equalizer only during the first iteration, while subsequent iterations use a conventional LMMSE equalizer. By providing LLR estimation close to that achieved by a MAP equalizer in the first iteration, we aim to positively impact the quality of soft information in subsequent iterations. This approach significantly reduces the average computational complexity per iteration, as the major computational effort is concentrated in the initial iteration.

\subsection{M-PAM Neural Network Equalizer for $M > 2$}
Hitherto, our NN-based models were designed to perform equalization on two-level sent signals, where \(x_n \in \{-1, +1\}\). We now extend our method to support M-PAM with \(M>2\). Let \(q = \log_2 M\); each symbol $x_n$ is then defined by a \(q\)-tuple of code bits $\mathbf{c}_n=(c_{n}^0,\ldots,c_{n}^{m}\ldots,c_{n}^{q-1})$, mapped using a binary Gray function.
Each code bit in \(\mathbf{c}_n\) continues to admit two levels, allowing us to utilize our existing EqzNet models in a per-bit equalization strategy. To this end, for the $m$-th bit in the $q$-tuple, we maintain a designated variation of EqzNet that takes as an input observations frame $\mathbf{z}_n$ and outputs $f_{\boldsymbol{\theta}_{m}}(\mathbf{z}_n)$, corresponding to estimated LLR of that bit, $L(c_{n}^{m})$. During training, we adjust the learnable parameter set $\boldsymbol{\theta}_{m}$ of the EqzNet of each code bit to minimize the loss function at  (\ref{eq:loss}). In Section \ref{sec:experiments and results}, we test our equalizers on 4-PAM signals.

\begin{figure*}[htb]
    \centering 
\begin{subfigure}{0.3\textwidth}
  \includegraphics[width=\linewidth]{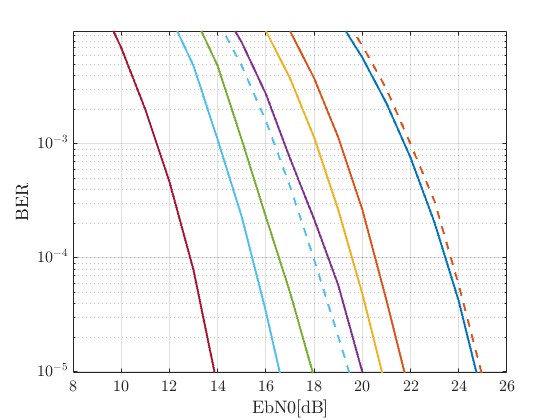}
  \caption{$\text{Uncoded 2-PAM over } \mathbf{h}_A$}
  \label{fig:ha_2_pam}
\end{subfigure}\hfil 
\begin{subfigure}{0.3\textwidth}
  \includegraphics[width=\linewidth]{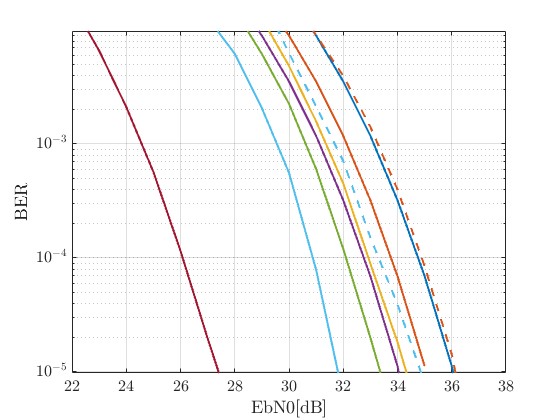}
  \caption{$\text{Uncoded 4-PAM over } \mathbf{h}_A$}
  \label{fig:ha_4_pam}
\end{subfigure}\hfil 
\begin{subfigure}{0.3\textwidth}
  \includegraphics[width=\linewidth]{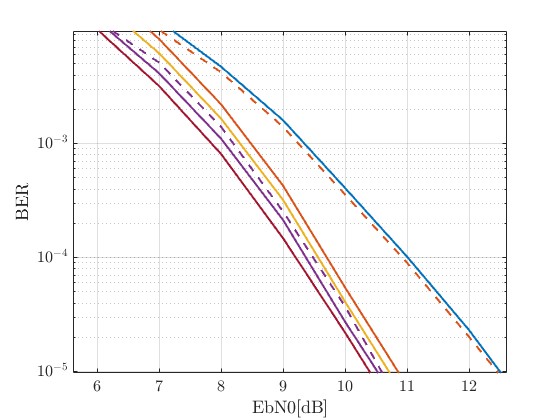}
  \caption{$\text{Uncoded 2-PAM over } \mathbf{h}_B$}
  \label{fig:hb_2_pam}
\end{subfigure}

\medskip
\begin{subfigure}{0.3\textwidth}
  \includegraphics[width=\linewidth]{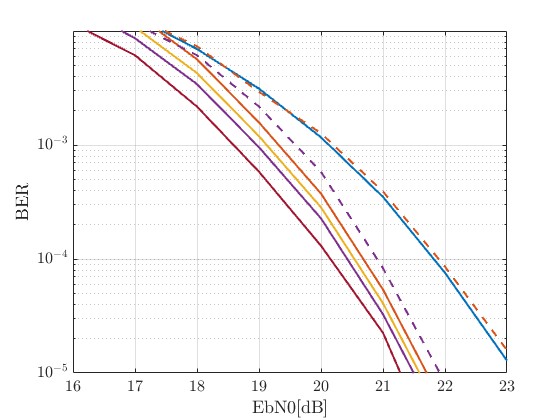}
  \caption{$\text{Uncoded 4-PAM over } \mathbf{h}_B$}
  \label{fig:hb_4_pam}
\end{subfigure}\hfil 
\begin{subfigure}{0.3\textwidth}
  \includegraphics[width=\linewidth]{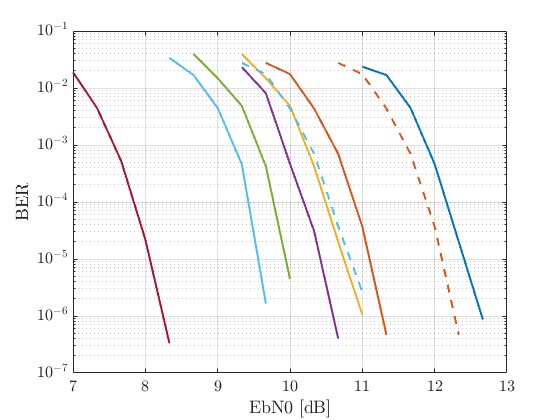}
  \caption{$\text{Turbo 2-PAM over } \mathbf{h}_A$}
  \label{fig:ha_2_pam_turbo}
\end{subfigure}\hfil 
\begin{subfigure}{0.3\textwidth}
  \includegraphics[width=\linewidth]{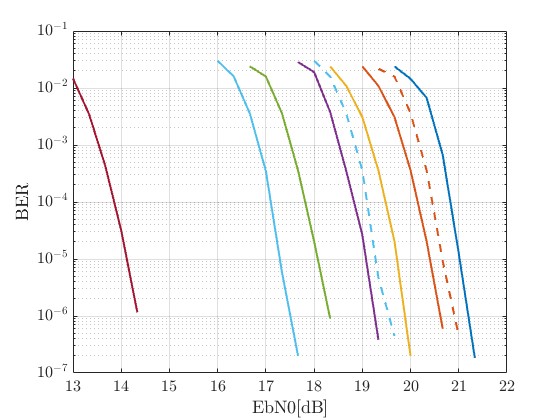}
  \caption{$\text{Turbo 4-PAM over } \mathbf{h}_A$}
  \label{fig:ha_4_pam_turbo}
\end{subfigure}
\caption{BER performance. 
\raisebox{0.5ex}{\textcolor{matlabblue}{\rule{1em}{1pt}}} LMMSE, 
\raisebox{0.5ex}{\textcolor{matorange}{\rule{1em}{1pt}}} $2$-EqzNet, 
\raisebox{0.5ex}{\textcolor{matyellow}{\rule{1em}{1pt}}}{$4$-EqzNet}, 
\raisebox{0.5ex}{\textcolor{matpurple}{\rule{1em}{1pt}}}{$6$-EqzNet}, 
\raisebox{0.5ex}{\textcolor{matgreen}{\rule{1em}{1pt}}}{($6, 2$)-EqzNet},
\raisebox{0.5ex}{\textcolor{matlightblue}{\rule{1em}{1pt}}}{($8, 2$)-EqzNet} and 
\raisebox{0.5ex}{\textcolor{matred}{\rule{1em}{1pt}}}{MAP}.
 Dashed-line curves correspond to similar architecture with random initialization.}
\label{fig:ber_performance}
\vspace{-1.em}
\end{figure*}

\section{Computational Complexity Analysis}
\label{sec:complexity}
The first hidden-layer is the most parameter-dense layer in $K$-EqzNet or ($K,2$)-EqzNet models. Each neuron in this layer contains learnable parameters equal to the length of the LMMSE filter plus a bias term. The second hidden-layer, if present, is narrower and contributes minimally to the overall computational burden. Let $K$ represent the number of neurons at the first hidden-layer, and let $\mathcal{O}(\text{LMMSE})$ denote the complexity of applying the LMMSE filter on observations frame $\mathbf{z}_n$. Thus, the total computational complexity of the network can be approximated as $K \cdot \mathcal{O}(\text{LMMSE})$ per bit, up to a constant factor. For $K<10$, our models are computationally more efficient compared to the MAP approach, such as the full-state BCJR algorithm, whose complexity is $\mathcal{O}(|\mathcal{B}|^{M_h})$.

\begin{figure*}[ht] 
    \centering
    \begin{minipage}[t]{0.225\textwidth} 
        \centering
        \includegraphics[width=\linewidth]{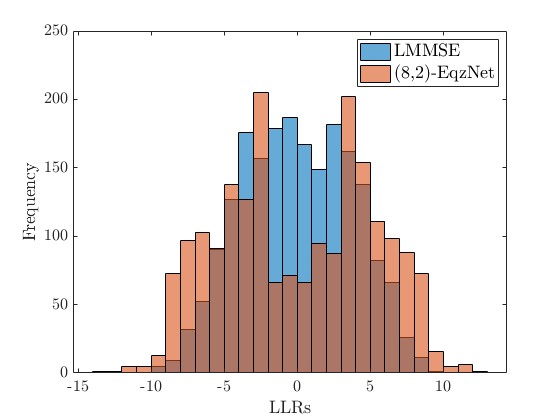}
        \subcaption{$E_b/N_0$ at 14dB}
        \label{fig:figure1}
    \end{minipage}
    \begin{minipage}[t]{0.225\textwidth}
        \centering
        \includegraphics[width=\linewidth]{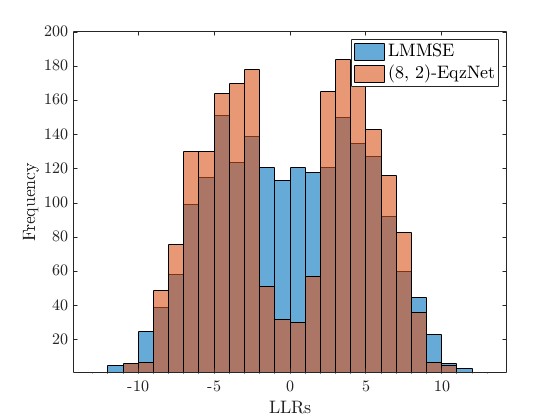}
        \subcaption{$E_b/N_0$ at 16dB}
        \label{fig:figure2}
    \end{minipage}
    \begin{minipage}[t]{0.225\textwidth}
        \centering
        \includegraphics[width=\linewidth]{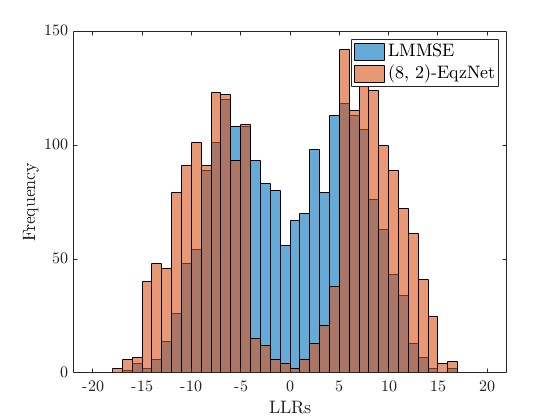}
        \subcaption{$E_b/N_0$ at 21dB}
        \label{fig:figure3}
    \end{minipage}
    \caption{LLRs from ($8,2$)-EqzNet and LMMSE for 2-PAM signals transmitted over the $\mathbf{h}_A$ channel at various SNR levels.}
    \label{fig:LLR_distributions}
    \vspace{-1.5em}
\end{figure*}

\section{Experiments and Result}
\label{sec:experiments and results}
In this section, we present several architectures for various EqzNet models, as outlined in Section \ref{sec:proposed_nn}, designed to operate on ISI channels with memory $M_h=4$. Primarily, our focus lies on the $K$-EqzNet (Fig. \ref{fig:multi_neurons}) and ($K+L$, 2)-EqzNet (Fig. \ref{fig:finaml_model}). While there are infinitely many combinations of $K$ and $L$ to choose from, we have selected a subset of EqzNet models with increasing complexity, yet ensuring that the overall complexity remains below $10\cdot\mathcal{O}(LMMSE)$.
For $K$-EqzNet and $M_h=4$, we consider $K\in\{2, 4, 6\}$, meaning that $K$ is increasing until $K=M_h+2$. As for the ($K+L,2$)-EqzNet model, we take $K\in\{4, 6\}$ and $L=2$. Models with greater complexity or different combinations of $K$ and $L$ yielded only marginal gains and were not pursued further.
We then compare the BER performance of various EqzNet models to that of MAP, LMMSE, and EqzNet models randomly initialized using standard normal distribution.

To determine the optimal input frame length for each channel and $M$-PAM modulation scheme across both the LMMSE and EqzNets, we execute the BCJR algorithm using a sliding window of different sizes. This process continues until the BER performance aligns with that of the full-state BCJR algorithm.

Throughout the simulations we use 2-PAM and 4-PAM signals and two unit power ISI channels: 
\begin{align*}
    \mathbf{h}_{A} &=[0.135,0.450,0.750,0.450,0.135] \\
    \mathbf{h}_B &= [0.877, 0.438, 0.168, 0.084, 0.059],
\end{align*}
Channel $\mathbf{h}_{A}$ imposes significant ISI, while $\mathbf{h}_{B}$ imposes milder ISI.
We test BER performance in uncoded and turbo equalization modes. For the latter, the information bits are protected using linear code LDPC($1998, 1776$). 
We train an EqzNet model for every channel per M-PAM and SNR \(E_b/N_0\), utilizing $10^6$ UID information bits for training and $10^{8}$ for testing.
As an optimization method Adam \cite{b14} was chosen. In all cases, the performance at a BER of $10^{-5}$ is compared.

\begin{figure}[t]
  \begin{subfigure}{0.49\columnwidth}
  \includegraphics[width=\textwidth]{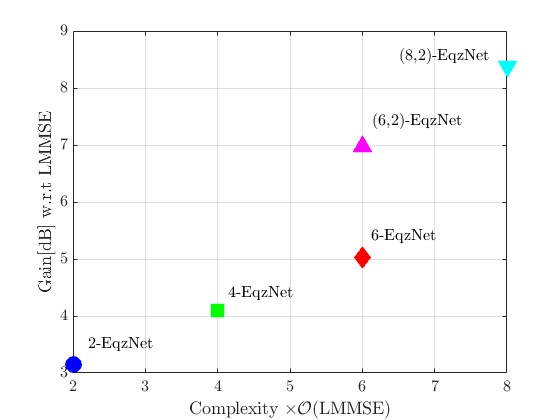}
  \caption{2-PAM signals}
  \end{subfigure}
  \begin{subfigure}{0.49\columnwidth}
  \includegraphics[width=\textwidth]{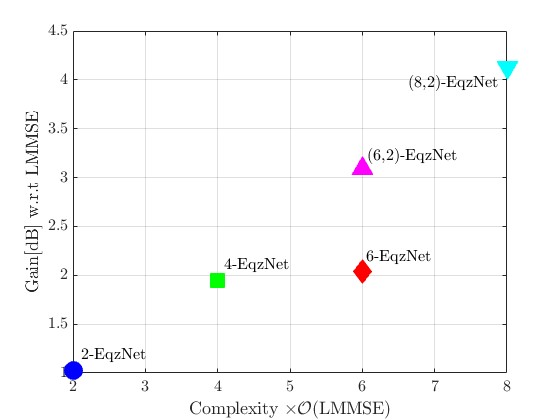}
  \caption{4-PAM signals} 
  \end{subfigure} 
  \caption{EqzNet Gain Vs. Complexity for channel $\mathbf{h}_A$ is active.}
  \label{fig:factor_gain}
  \vspace{-1.5em}
\end{figure}

Fig. \ref{fig:ha_2_pam}-\ref{fig:hb_4_pam} illustrate the BER performance for uncoded equalization of 2-PAM and 4-PAM signals over our studied channels. Notably, as small as $2$-EqzNet model, at $2\cdot\mathcal{O}(LMMSE)$ complexity, outperforms LMMSE when initialized with our method. As complexity increases, our EqzNet models consistently achieve greater gains over LMMSE. Specifically, for 2-PAM signals sent over severe ISI channel $\mathbf{h}_A$, the $2$-EqzNet achieves about 3dB gain while (8,2)-EqzNet achieves 8.4dB.
At 4-PAM sent over $\mathbf{h}_A$, we exhibit 1.25dB and 4.1dB with $2$-EqzNet and ($8,2$)-EqzNet, respectively. The gains of all EqzNet models against complexity when channel $\mathbf{h}_A$ is present are summarized in Fig. \ref{fig:factor_gain}.
When $6$-EqzNet is applied to $\mathbf{h}_B$, it nearly matches the performance of the MAP equalizer for both modulation. This demonstrates that a low-complexity model, when initialized with our method, can significantly enhance the gain over LMMSE with only a modest increase in complexity. Remarkably, for randomly initialized $2$-EqzNet models, the results mostly indicate negligible gains over LMMSE. Moreover, in Fig. \ref{fig:ha_2_pam} a gap of approximately 3.5dB was observed between the initialized and uninitialized versions of the ($8,2$)-EqzNet.
This highlights our assertion that initialization aids in navigating around poor local minima.

Fig. \ref{fig:ha_2_pam_turbo}-\ref{fig:ha_4_pam_turbo} demonstrate BER performance of turbo equalization when channel $\mathbf{h}_A$ is present. Here, EqzNet is used only in the first iteration, with LMMSE applied in the subsequent iterations. As expected, the gap between MAP and the rest of the equalizers, including LMMSE, diminishes. Evidently, our EqzNet models still outperform LMMSE, which reinforces the assumption that the quality of output LLRs play key role in turbo equalization. Fig. \ref{fig:LLR_distributions} offers additional insight by comparing the LLRs from the ($8,2$)-EqzNet to those from LMMSE. Evidently, ($8,2$)-EqzNet produces LLR distributions with higher magnitude and fewer LLRs around zero, implying greater confidence in bit decisions.

\section{Conclusion}
This paper focuses on developing NN-based equalizers characterized by a small number of parameters and low-complexity. We have demonstrated that optimizing NNs with a limited parameter set results in convergence at a local minimum with marginal performance enhancement compared to LMMSE. To this end, we introduced a FC-NN equalizer comprising a single hidden-layer initialized with shifted LMMSE taps $\mathbf{f}_k$ to estimated the ISI induced by adjacent symbols. This NN-based equalizer resolves optimization challenges and enhances LMMSE performance. Additionally, we proposed two architectures building upon this foundational hidden-layer design, incrementally improving performance with a linear increase in complexity. Our proposed FC-NN is trained to generate LLRs close resembling those of a MAP decoder. By employing FC-NN LLRs exclusively in the initial turbo equalization iteration, and subsequently switching to LMMSE LLRs for subsequent iterations, we achieved a notable performance boost over turbo equalization with LMMSE. This study asserts the potential of low-complexity NN equalizers, offering substantial performance gains over LMMSE with modest computational overhead.

\vspace{12pt}

\end{document}